\documentstyle[12pt,epsf]{article}
\newcommand\pubnumber{SLAC-PUB-8002}
\newcommand\pubdate{November, 1998}

\def\Title#1{\begin{center} {\Large #1 } \end{center}}
\def\Author#1{\begin{center}{ \sc #1} \end{center}}
\def\Address#1{\begin{center}{ \it #1} \end{center}}

\def\submit#1{\begin{center}Submitted to {\sl #1} \end{center}}
\def\doeack{\footnote{Work supported by the Department of Energy,
                     contract DE--AC03--76SF00515.}}
\def\SLAC{Stanford Linear Accelerator Center\\
    Stanford University, Stanford, California 94309 USA}
\newcommand\pubblock{\rightline{\begin{tabular}{l} \pubnumber\\
         \pubdate  \end{tabular}}}
\newenvironment{Abstract}{\begin{quotation} \begin{center}
                       ABSTRACT
     \end{center}\bigskip  }{\end{quotation}}
\def\beq{\begin{equation}}
\def\eeq#1{\label{#1}\end{equation}}
\def\eeqn{\end{equation}}
\def\beqa{\begin{eqnarray}}
\def\eeqa#1{\label{#1}\end{eqnarray}}
\def\eeqan{\end{eqnarray}}
\def\CR{\nonumber \\ }
\def\leqn#1{(\ref{#1})}

\def\Journal#1#2#3#4{{#1} {\bf #2}, #3 (#4)}

\def\NPB{{\em Nucl. Phys.} B}
\def\PLB{{\em Phys. Lett.}  B}

\def\PRD{{\em Phys. Rev.} D}

\def\JETP{{\em Sov. Phys. JETP}}

\let\bar=\overbar
\def\Dslash{\not{\hbox{\kern-4pt $D$}}}
\def\dslash{\not{\hbox{\kern-2pt $\del$}}}
\def\half{\frac{1}{2}}

\def\L{{\cal L}}

\def\del{\partial}

\def\Pl{{\mbox{\scriptsize Pl}}}

\def\ee{e^+e^-}

\def\msb{{\bar{\ssstyle M \kern -1pt S}}}

\def\etal{{\it et al.}}

\begin{document}
\begin{titlepage}
\pubblock

\vfill
\Title{Collider Signatures of New Large Space Dimensions}
\vfill
\Author{Eugene A. Mirabelli, Maxim Perelstein, and Michael E. Peskin\doeack}
\Address{\SLAC}
\vfill
\begin{Abstract}
Recently, Arkani-Hamed, Dimopoulos, and Dvali have proposed that there are
extra compact dimensions of space, accessible to gravity but not to ordinary
matter, which could be macroscopically large.  In this letter, we argue that
high-energy collider processes in which gravitons are radiated into these
new dimensions place significant, model-independent constraints on this
picture.  We present the constraints from anomalous single photon production
at $\ee$ colliders and from monojet production at hadron colliders. 
\end{Abstract}
\medskip
\submit{Physical Review Letters}

\vfill
\end{titlepage}
\def\thefootnote{\fnsymbol{footnote}}
\setcounter{footnote}{0}

The Standard Model of strong, weak, and 
electromagnetic interactions has been dramatically successful in explaining
the rates of high-energy $\ee$ and $p\bar{p}$ reactions and
the properties of the $W$ and $Z$ bosons.  This great success, however,
has focused attention on the fact the Standard Model (SM) requires a number
of choices for its input parameters which are very difficult to understand.
Among these are the value of the Higgs boson mass parameter $\mu^2$
and the value
of the cosmological constant $\lambda$.  If one assumes that the most 
fundamental scale in Nature is the Planck scale, $M_\Pl = G_N^{-1/2} =
10^{19}$ GeV and 
writes these parameters in terms of this scale, one finds 
$\mu^2 \sim 10^{-34} M_\Pl^2$, $\lambda \sim 10^{-116} M_\Pl^4$.

The mystery of these small parameters has motivated many authors to consider
radical ideas for the manner in which gravity is unified with the other
fundamental interactions.  The introduction of supersymmetry can lower the 
natural mass scale for $\mu^2$ and $\lambda$ to 1 TeV.  This ameliorates the
problem of the Higgs mass but is not nearly enough of a reduction to solve the 
cosmological constant problem. Many authors have investigated whether a string
theory of quantum gravity can provide a further reduction.  String theory
includes the possibility of additional microscopic space dimensions.
In this context,
Antoniadis \cite{Anto} has proposed that Nature may contain additional 
compact dimensions of size  $\hbar$/TeV \cite{Lykken}.

Recently, several groups \cite{Sund, ADD, ST}
have extended this proposal using new ideas about
the strong-coupling behavior of string theory.  In this regime, string theory
may contain solitons or mirror surfaces that occupy lower-dimensional 
hypersurfaces,  with some species of particle restricted to these objects.
  One 
can then imagine that the quarks, leptons, and gauge bosons of the SM live
on a 4-dimensional hypersurface inside
 the full space-time, while gravity lives in the
full, higher-dimensional space. Arkani-Hamed, Dimopoulos, 
and Dvali (ADD) \cite{ADD} have argued that, in these models, the fundamental
gravitational scale can be as low as TeV energies, while the size of the 
extra dimensions can be as large as a millimeter.  

If indeed gravity becomes strong at TeV energies, gravitons 
should be radiated
at significant rates in high-energy particle collisions. In collider
experiments, higher-dimensional gravitons (G) appear as massive 
spin-2 neutral particles which are not observed by collider detectors.
As ADD pointed out, G radiation leads to missing-energy signatures in
 which a photon
or a jet is produced with no observable particle balancing its transverse
momentum.  In this paper, we compute the rates of the missing-energy 
processes
\beq
  \ee\to \gamma + (\hbox{missing}) \ , \qquad p\bar p \to \hbox{jet} + 
 (\hbox{missing}) 
\eeq{processes}
and the corresponding experimental constraints.
  We show that these experiments actually give the strongest
present constraints on the size of the extra dimensions, and that future 
experiments will have even better sensitivity.

{\bf Conventions.}  In this paper, we assume that the 
gravitational field is the only field that propagates in 
 the extra dimensions.
It is likely that, in realistic models, the extra dimensions will also 
contain scalar, vector, and even fermion fields that couple to the 
SM particles with gravitational strength.  These particles would produce
additional, model-dependent, missing-energy signatures beyond those
we consider here.  We also assume that the typical momenta with which 
gravitons are emitted are small relative to the thickness of the hypersurface
on which the SM particles live, and also relative  to the fundamental
gravitational scale.    In this limit, the higher-dimensional
gravitational field couples to the energy-momentum tensor of the SM, with
precisely the coupling of standard 4-dimensional gravity \cite{SundA}.
 To compute the rate of 
emission of a single G particle, we interpret the G  momentum in
 the extra
dimensions as a 4-dimensional mass for this spin-2 particle
and use the Lagrangian
\beq
    \delta \L = - (8\pi G_N)^{1/2}  G_{\mu\nu} T^{\mu\nu} \ .
\eeq{basicL}
  Under our assumptions,
this coupling is model-independent.

For definiteness, we will assume that there are $n$ extra dimensions,
and that these are compactified on an $n$-dimensional
torus of periodicity $2\pi R$.   With our choices, the 
gravitational potential for 
 $r$ just larger than  $R$ takes the form
\beq
    V(r) = - {G_N m_1 m_2\over r}\left(1 + 2n \, e^{-r/R} + \cdots \right) \ .
\eeq{longV}
Macroscopic measurements of the force of gravity constrain $R$ to be less
than a millimeter; for example, for $n=2$,  $R < 0.77$ mm 
 at 95\% confidence \cite{Mitro,Price}.  For $r \ll R$, the potential goes
over to the $r^{-(n+1)}$ dependence characteristic of the higher-dimensional
space.  The coefficient of this potential is proportional to 
 a power of fundamental 
gravitational scale.  ADD define this scale by the formula
\beq
       M^{n+2} R^n = (4\pi G_N)^{-1}  \ .
\eeq{Mdefin}
With this definition, $M = 1$ TeV
corresponds to $R = 0.68$ mm for $n=2$ and to $R = 3.0 \times 10^{-12}$ cm
 for $n=6$.

In computing the emission of G particles, it 
is necessary to sum  over the possible values of the 
higher-dimensional
momenta $k_\perp$. This is,  equivalently, a sum or integral over values of 
the 4-dimensional mass
\beq
   \sum_{k_\perp} =     R^n  \int d^n m = 
  \half \Omega_n R^n \,  \int\,  (m^2)^{(n-2)/2} d\,  m^2 =  
    {\Omega_n\over 8 \pi} M^{-(n+2)}  \int\,  (m^2)^{(n-2)/2} d\,  m^2 \,
         G_N^{-1} \ , 
\eeq{mps}
where $\Omega_n$ is the volume of the unit sphere in $n$ dimensions (= $2\pi$
for $n=2$).

{\bf Electron-positron collisions.}  With these conventions, it is now
straightforward to compute the rate of $\ee$ annihilation into an 
anomalous single photon recoiling against an unobserved G.  This reaction
could potentially be observed at the CERN $\ee$ collider LEP~2, or at 
a higher-energy $\ee$ collider.

The differential cross section for the 
reaction $e^-_Le^+_R\to \gamma G$, considered 
 in the center of mass system for a G of
mass $m$,  is given by \cite{helic}
\beqa
{d\sigma\over d\cos\theta}  &=& {\pi \alpha G_N \over 1-m^2/s}\Biggl[(1 + 
   \cos^2\theta )\left(1 + \bigl({m^2\over s}\bigr)^4\right) \CR
     & & \hskip 0.5in  + \left( { 1 - 3 \cos^2\theta + 4 \cos^4\theta\over
          1- \cos^2\theta}\right){m^2\over s }\left(1 +  
           \bigl({m^2\over s}\bigr)^2\right) 
     + 6 \cos^2\theta \bigl({m^2\over s}\bigr)^2
                                          \Biggr] \ .
\eeqa{eeformula}
The same formula holds for $e^-_R e^+_L$; the helicity-violating 
cross sections are zero.  These expressions
 must be integrated over the phase space \leqn{mps}.  The
cross section behaves as $\sigma \sim  s^{n/2}/M^{n+2}$.  Thus, the 
production of anomalous single photons increases dramatically as the 
center-of-mass energy is raised.

In the SM, single photon events are produced in the reaction $\ee \to 
\gamma \nu \bar \nu$, which can proceed through $s$-channel $Z^0$ exchange
or (for the case of $\nu_e$) through $t$-channel $W$ exchange \cite{Berends}.
The effect of $G$ emission would be observable as an enhancement of the 
cross section for single-$\gamma$ production above that of this SM source.
The single-$\gamma$ cross section has been measured by the  
 LEP 2 experimental groups at
$\sqrt{s}$ = 183 GeV \cite{LEP}. 
  The measurements agree with the SM  prediction to 6\% 
accuracy.  If we integrate our prediction for the G signal over the 
kinematic region studied in these experiments, we find, for the case 
$n=2$, the limits $R < 0.48$ mm, $M > 1200$ GeV at 95\% confidence.  Limits
for higher values of $n$ are given in Table \ref{thetable}.

%%%%%%%%%%%%%%%%%%%%%%%%%%%%%%%%%%%%%%%%%%%%%%%%%%%%%%%%%%%%%%%%%%%%%%
\begin{figure}[t]
\begin{center}
\leavevmode
{\epsfysize=3.00truein \epsfbox{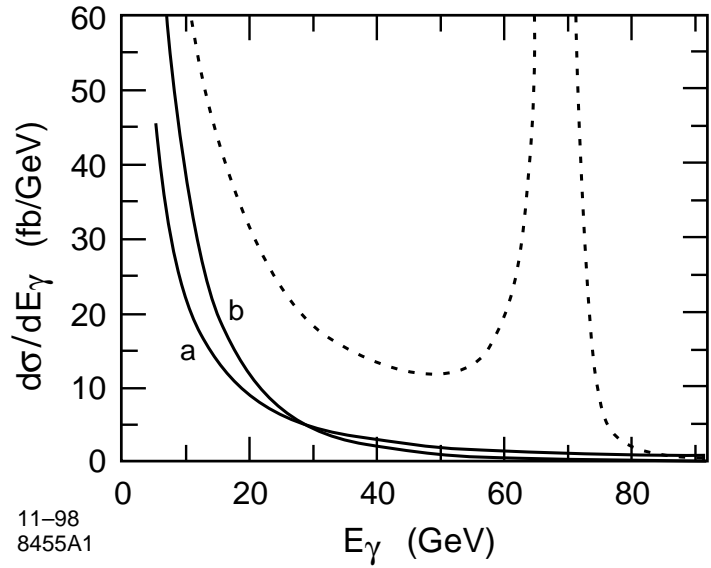}}
\end{center}
 \caption{Energy spectrum of single photons recoiling against 
    higher-dimensional gravitons G,
   computed for $\ee$ collisions at 
  $\sqrt{s} = 183$ GeV with an angular cut
    $|\cos\theta| < 0.95$.  The dotted curve is the Standard Model
 expectation.
 The solid curves show the additional cross section expected in the model of
      ref. \cite{ADD} with  (a) $n=2$, $M = 1200$ GeV, (b)
              $n=6$, $M = 520$ GeV.}
\label{eefig}
\end{figure}
%%%%%%%%%%%%%%%%%%%%%%%%%%%%%%%%%%%%%%%%%%%%%%%%%%%%%%%%%%%%%%%%%%%%%%

In 
Figure~\ref{eefig}, we show the energy distribution of single photons
recoiling against G particles 
for the cases $n=2$ and $n=6$, for the parameter values at our limit, 
compared to the single-photon distribution from the SM process.
The peak in the SM cross section results from the process in which 
the $\gamma$ recoils against an on-shell $Z^0$ which decays invisibly.
Some additional advantage can be gained, then, in applying a cut which
excludes this peak.  For the kinematic region $20 < E_\gamma < 
50$ GeV, $|\cos\theta_\gamma| < 0.95$ and $\sqrt{s} = 183$ GeV, we
 find the cross section for G production
\beq
  \sigma =  267/M^4 \ , \ 19/M^6  \ , \ 0.65/M^8 \ \hbox{fb}\ ,
\eeq{LEPcross}
for $n=2,4,6$ and  $M$ in TeV.

Higher-energy studies of 
 $\ee$ annihilation will be done at a linear $\ee$
collider (LC).  We have already noted that higher energy alone should lead
 to much higher sensitivity
to G production.  But the  LC also offers another advantage, the 
 possibility of electron beam 
polarization, which can be used to suppress the dominant $t$-channel $W$
exchange piece of the SM background process.  At $\sqrt{s} = 1$ TeV, with
electron polarization $P = +0.9$ (right-handed), integrating over
the kinematic region 
$50$ GeV $< E_\gamma < 400$ GeV, $|\cos\theta_\gamma| < 0.95$, we find 
a SM background cross section of 82 fb and a G signal cross section
of 
\beq
  \sigma =  20/M^4 \ , \ 46/M^6  \ , \ 55/M^8 \ \hbox{pb}\ ,
\eeq{LCcross}
for $n=2,4,6$ and  $M$ in TeV.
To quantify the effect of this measurement, we assume 
that this cross section can be measured with 5\% accuracy,
 and that the value to be found agrees with the SM.  Then
the measurement would give very strong limits on $R$ and $M$ which are listed
in Table~\ref{thetable}.

{\bf Proton-antiproton collisions.}   In a similar way,  proton-antiproton
collisions can lead to processes in which a single parton is produced
at large transverse momentum 
recoiling against a G particle.   This leads to a monojet
 signature of 
G production---a jet plus missing transverse energy ($E_T$)---which
 may be visible at the Fermilab Tevatron collider.
The search for this reaction complements the search in $\ee$ reactions in
the familiar way, with the higher energy available in hadron collisions
compensating important losses in the definition of the signal.

The production of jets with large $E_T$ recoiling against G particles can 
arise from the 
parton subprocesses  $q\bar q \to G g$, $q g \to q G$,
 $\bar q g \to \bar q G$, and $gg \to g G$.
The polarization- and color-averaged cross section for $q\bar q \to gG$
can be obtained directly from Eq. \leqn{eeformula}
\beqa
{d\sigma\over d \cos\theta}  &=& {2\over 9}
 {\pi \alpha_s G_N \over 1-m^2/s}\Biggl[(2 - {4ut\over (s-m^2)^2}) 
  \left(1 + \bigl({m^2\over s}\bigr)^4\right) \CR
     & & \hskip -0.4 in  + \left(2 { (s-m^2)^2\over 4 ut} - 5 
      + 4 {4ut\over (s-m^2)^2}\right){m^2\over s }\left(1 +  
   \bigl({m^2\over s}\bigr)^2\right) + 6 \left( {u-t\over s-m^2}\right)^2
                          \bigl({m^2\over s}\bigr)^2
                                          \Biggr] \ ,
\eeqa{qqformula}
where $s,t,u$ are the Mandelstam variables: $t,u = -\half s (1-m^2/s)(1\mp 
\cos \theta)$.
The cross section for $qg \to q G$ 
 can be obtained from this expression 
by crossing $s \leftrightarrow t$:
\beqa
{d\sigma\over d \cos\theta}  &=& 
 {\pi \alpha_s G_N (-t/s)(1-m^2/s) \over 12\,(1-m^2/t)^2}
 \Biggl[(2 - {4us\over (t-m^2)^2}) 
  \left(1 + \bigl({m^2\over t}\bigr)^4\right) \CR
     & & \hskip -0.4 in  + \left(2 { (t-m^2)^2\over 4 us} - 5 
      + 4 {4us\over (t-m^2)^2}\right){m^2\over t }\left(1 +  
   \bigl({m^2\over t}\bigr)^2\right) + 6 \left( {s-u\over t-m^2}\right)^2
                          \bigl({m^2\over t}\bigr)^2
                                          \Biggr] \ .
\eeqa{qgformula}
The cross section for $\bar q g \to \bar q G$ is also
 given by \leqn{qgformula}.
  For the process
 $gg \to g G$, we find the polarization- and color-averaged cross 
section \cite{helictwo}
\beqa
{d\sigma\over d \cos\theta }  &=& {3\over 16}
      {\pi \alpha_s G_N \over (1-m^2/s)(1-\cos^2\theta)}\Biggl[(3 + 
   \cos^2\theta )^2\left(1 + \bigl({m^2\over s}\bigr)^4\right) \CR
     & & \hskip 0.1in  - 4\, ( 7 + \cos^4\theta){m^2\over s }\left(1 +  
           \bigl({m^2\over s}\bigr)^2\right) 
          + 6\, (9 - 2  \cos^2\theta +  \cos^4\theta)  
                      \bigl({m^2\over s}\bigr)^2  
                                          \Biggr] \ .
\eeqa{Ggformula}
All of these formulae must be integrated over the G mass spectrum
using the measure \leqn{mps}.
The rate of monojet production can then be found by integrating these 
cross sections with appropriate parton distributions.

The processes $q\bar q \to g Z^0$, $q g \to q Z^0$, followed by an invisible
decay of the $Z^0$, give an irreducible physics background to G production.
We will refer to this process as the `SM background', and we will estimate
the observability of our signal by comparing its cross section to that of
this reaction.  There are other important background sources from 
mismeasured jets and $W$ production with forward leptons, but these 
backgrounds decrease sharply as the lower bound on missing $E_T$ is 
increased.
Unlike the case of $\ee$ reactions, the detector does not measure the 
imbalance in longitudinal momentum, and there is not enough kinematic
information from the single observed
jet to exclude the kinematic region in which
the $Z^0$ is on-shell.  On the other hand, the parton center of mass 
energies available at the Tevatron are higher than those of LEP 2, and we
have seen that the G signal increases rapidly with energy.  It is therefore
reasonable to look for the monojet signal as an excess above
 the SM cross section
for on-shell
$Z^0$ production.  Though it is not so easy to compute the SM background
 rate accurately, this rate
 can be normalized to the corresponding process in 
which the $Z^0$ decays to a lepton pair.

%%%%%%%%%%%%%%%%%%%%%%%%%%%%%%%%%%%%%%%%%%%%%%%%%%%%%%%%%%%%%%%%%%%%%%
\begin{figure}[t]
\begin{center}
\leavevmode
{\epsfysize=3.00truein \epsfbox{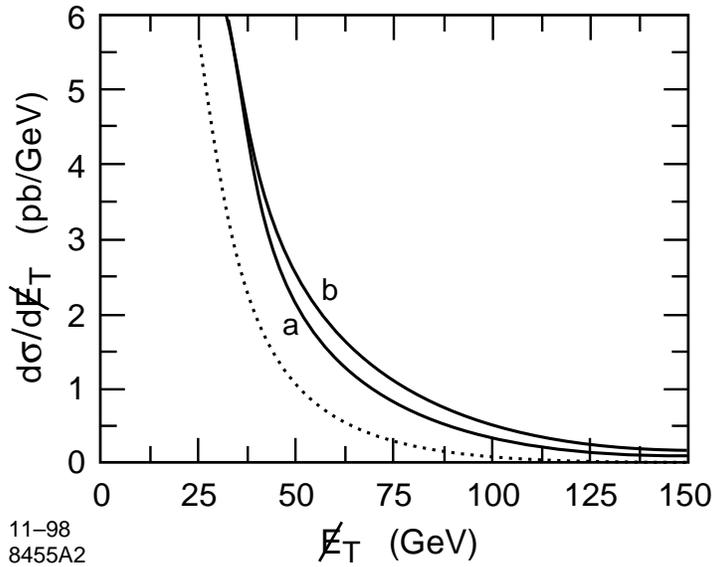}}
\end{center}
 \caption{Spectrum of missing energy in events with one jet,
    computed 
   for $p\bar p $ collisions at  $\sqrt{s} = 1.8$ TeV, with a rapidity cut
    $|y| < 2.4$.  The dotted curve is the Standard Model
 expectation.
 The solid curves show the additional cross section expected in the model of
      ref. \cite{ADD} with  (a) $n=2$, $M = 750$ GeV, (b) $n=6$,
 $M = 610$ GeV.}
\label{ppfig}
\end{figure}
%%%%%%%%%%%%%%%%%%%%%%%%%%%%%%%%%%%%%%%%%%%%%%%%%%%%%%%%%%%%%%%%%%%%%%

The CDF collaboration has presented a bound on monojet production based on
its first 4.7 pb$^{-1}$ of data in $p\bar p$ collisions 
 at $\sqrt{s} = 1.8$ TeV \cite{Markel,later}.  This analysis searched for
 events with missing $E_T$ greater than 30 GeV
and one jet in the rapidity region $|y| < 1.2$.   The result was consistent
with the $Z^0$ background and can be represented as a limit on the number of
neutrinos into which the $Z^0$ decays:  $N_\nu < 5.0$ (95\% confidence).
We convert this to a limit on G production by comparing the 
cross sections for the G signal
and the SM process, computed in the same framework.  For simplicity, we 
carry out the calculations of both signal and background at the leading
order in QCD, using the CTEQ4 lowest-order (set 3) structure functions
\cite{CTEQ}.  We find a SM background  cross section of 37 pb for the cuts
 listed
above, and, for $n=2$, a signal cross section of 20 pb/$M^4$.  This implies
a limit $R < 1.2$ mm, $M > 750$ GeV.   Limits
for higher values of $n$ are given in Table \ref{thetable}.  In 
Figure~\ref{ppfig}, we show the missing $E_T$ spectrum of the signal and 
background processes.

It is advantageous to make a tighter cut on missing $E_T$ to remove the
backgrounds from mismeasured jets which were a problem for the CDF analysis
\cite{Markel}.
Integrating the signal and background rates over the region with 
missing $E_T > 60$ GeV and jet rapidity $|y| < 2.4$, we find a SM 
background cross section  for Z production of 10 pb, and signal 
cross sections in the ratios
\beq
   S/B =  0.85/M^4 \ , \  0.15/M^6 \ , \ 0.052/M^8 \ ,
\eeq{newTeV} 
for $n=2,4,6$ and  $M$ in TeV.
Assuming that this measurement can be performed with 20\% accuracy, and 
 that the value to be found agrees with the SM,  we find the potential limits
 on $R$ and $M$ listed in the third line of Table~\ref{thetable}.
 
Hadron-hadron collisions will be studied at higher energy at the 
CERN LHC.  At the LHC, most collisions are between gluons, since the 
gluon structure functions rise rapidly at low $x$.  This suppresses the
SM contribution, since gluon-gluon collisions cannot lead to $Z^0$ 
production at the leader order in $\alpha_s$.
 However, we find that the most important 
contributions to G production also involve quarks, since the 
enhancement of the cross section at high energy partially
 compensates the falloff
of the structure functions.  
  Repeating the analysis  leading to \leqn{newTeV} at the LHC
energy of 14 TeV using the kinematic cuts $E_T > 200$ GeV, $|y| < 5$,
we find a SM background cross section of 11 pb and signal cross sections
in the ratios 
\beq
   S/B =  110/M^4 \ , \  420/M^6 \ , \ 3600/M^8 \ ,
\eeq{LHC} 
for $n=2,4,6$ and  $M$ in TeV.  With the same assumptions as for \leqn{newTeV},
we find the potential limits listed in the last line of Table~\ref{thetable}.
It is important to note that, in the case $n=6$, the dominant parton-parton
center of mass energies are comparable to the quoted limit on $M$, so
the effective coupling \leqn{basicL} might not be appropriate for this case.

%%%%%%%%%%%%%%%%%%%%%%%%%%%%%%%%%%%%%%%%%%%%%%%%%%%%%%%%%%%%%%%%%%%%%
\begin{table}
\begin{center}
\begin{tabular}{l l | r | r | r } 
  Collider   & &  R / M  ($n=2$)  &  R / M  ($n=4$) &   R / M  ($n=6$) \\
           \hline\hline
  
   Present: &LEP 2   &   $4.8\times 10^{-2}$ / 1200
                                       & $1.9\times 10^{-9}$ / 730 & 
                              $6.9 \times 10^{-12}$ / 520 \\ \hline
 & Tevatron  &   $11.0 \times 10^{-2}$ / 750  & $2.4 \times 10^{-9}$ / 610 
              & $5.8 \times 10^{-12}$ / 610 \\ \hline \hline
  Future: &Tevatron &  $3.9 \times 10^{-2}$ / 1300 
            &  $1.4 \times 10^{-9}$ / 900
              & $4.0 \times 10^{-12}$ / 810     \\ \hline
&  LC & $1.2\times 10^{-3}$ / 7700  & $1.2\times 10^{-10}$ / 4500 & 
                              $6.5 \times 10^{-13}$ / 3100 \\ \hline
 & LHC &  $3.4\times 10^{-3}$ / 4500  & $1.9\times 10^{-10}$ / 3400 & 
                              $6.1 \times 10^{-13}$ / 3300 \\ \hline
\end{tabular}
\end{center}
 \caption{Current and future sensitivities to large extra dimensions,
expressed as 95\% confidence limits on the size of 
 extra dimensions $R$ (in cm) and the effective Planck scale $M$ (in GeV). 
 The assumptions
of each analysis are explained in the text.}
\label{thetable}
\end{table}
%%%%%%%%%%%%%%%%%%%%%%%%%%%%%%%%%%%%%%%%%%%%%%%%%%%%%%%%%%%%%%%%%%%%%%%%

{\bf Conclusions.}  In this paper, we have shown that high-energy collider
searches for events with missing energy and transverse momentum provide
a relevant, model-independent test of theories with large extra 
space dimensions.   Current high-energy experiments
at $\ee$ and $p\bar p$ colliders already place the strongest direct constraints
on these theories.
 Higher-energy experiments may place much stronger constraints.
Or,  more optimistically, they  may 
allow us to observe  an  excess of 
missing-energy events above the SM expectation, providing direct evidence
for this
remarkable extension of our conception of the universe.

{\bf Acknowledgements.} We are grateful to Nima Arkani-Hamed
 for suggesting this project and 
encouraging us along the way, and to Michael Barnett, 
Savas Dimopoulos, Lance Dixon, JoAnne Hewett, Ian Hinchliffe,
Teruki Kamon, Joseph Lykken, and Wayne Repko
for helpful discussions.  This work was supported by the Department of 
Energy under contract DE--AC03--76SF00515.
As this paper was being completed, we received a paper by  Giudice,
 Rattazzi, and Wells  that carries out an analysis very similar to the one
presented here \cite{WGR}.  Other high-energy physics limits on extra
dimensions have been discussed recently in \cite{extra}.

%%%%%%%%%%%%%%%%%%%%%%%%%%%%%%%%%%%%%%%%%%%%%%%%%%%%%%%%%%%%%%%%%%%%%%%

%%%%%%%%%%%%%%%%%%%%%%%%%%%%%%%%%%%%%%%%%%%%%%%%%%%%%%%%%%%%%%%%%%%%%%%%%


\begin{thebibliography}{99}
\bibitem{helic}
  As $m^2 \to 0$, this expression has a smooth limit in which only the
    helicity states $\gamma(-)G(+2)$ and $\gamma(+)G(-2)$ are produced.
\bibitem{helictwo}
  As $m^2 \to 0$, this expression has a smooth limit in which the only 
 nonzero 
    helicity amplitudes are those related by crossing to
      $g(+)g(+)g(-) \to G(+2)$ and $g(-)g(-)g(+) \to G(-2)$.
    Though it is not so obvious in this representation,
 the squared matrix element contributing
  to \leqn{Ggformula} is symmetric under permutations of $s$, $t$, and $u$.
\end{thebibliography}
\end{document}